\documentclass{jnmp}

\usepackage{amsmath}
\usepackage{cite}
\JNMPnumberwithin{equation}{section}

\theoremstyle{definition}

\theoremstyle{remark}

\numberwithin{equation}{section}
\newtheorem{step}{Step}[subsection]

\renewcommand{\vec}[1]{\mathbf{#1}}

\DeclareMathOperator{\rank}{rank}

\begin{document}

%  Headings
%
\renewcommand{\evenhead}{D Baldwin and W Hereman}
\renewcommand{\oddhead}{Symbolic Software for the Painlev\'e Test}

%  Titlepage
%
\thispagestyle{empty}

\FirstPageHead{13}{1}{2006}{\pageref{firstpage}--\pageref{lastpage}}{Article}
%  Parameters: Volume, number, year, page range, paper type
%  'Article' could be changed to 'Letter' or 'Review Article'

\copyrightnote{2006}{D Baldwin and W Hereman}

\Name{Symbolic Software for the Painlev\'e Test of 
Nonlinear Ordinary and Partial Differential Equations}

\label{firstpage}

\Author{Douglas BALDWIN~$^\dag$ and Willy HEREMAN~$^\dag$}

\Address{$^\dag$ Department of Mathematical and Computer Sciences, 
Colorado School of Mines, Golden, CO 80401, USA \\
~~E-mail: painlevetest@douglasbaldwin.com and  whereman@mines.edu}

\Date{Received April 22, 2005; Accepted in Revised Form June 5, 2005}

\begin{abstract}
\noindent
The automation of the traditional Painlev\'e test in 
\emph{Mathematica} is discussed.  
The package \texttt{PainleveTest.m} allows for the testing 
of polynomial systems of nonlinear ordinary and partial differential equations 
which may be parameterized by arbitrary functions (or constants).  
Except where limited by memory, there is no restriction on the 
number of independent or dependent variables.  
The package is quite robust in determining all the possible 
dominant behaviors of the Laurent series solutions of the differential 
equation.  
The omission of valid dominant behaviors is a common problem in 
many implementations of the Painlev\'e test, 
and these omissions often lead to erroneous results.  
Finally, our package is compared with the other available implementations 
of the Painlev\'e test.  
\end{abstract}

%  The paper
%
\section{Introduction}
\label{sec:intro}

Completely integrable nonlinear partial differential equations (PDEs) 
have remarkable properties, such as 
  infinitely many generalized symmetries, 
  infinitely many conservation laws, 
  the Painlev\'e property, 
  B\"ack\-lund and Dar\-boux transformations, 
  bilinear forms, and 
  Lax pairs 
  (cf.~\cite{Ablowitz80a,Fokas87,Lakshmanan83,Lamb80}).
Completely integrable equations model physically 
interesting wave phenomena in reaction-diffusion systems, 
population and molecular dynamics, nonlinear networks, chemical reactions, 
and waves in material science. 
%(in particular solid mechanics and elastic materials).  
By investigating the complete integrability of a nonlinear PDE, 
one gains important insight into the structure of the equation 
and the nature of its solutions.  

Broadly speaking, Painlev\'e analysis is the study of the singularity 
structure of differential equations.  
%Painlev\'e analysis is concerned with how the singularities of the solutions 
%depend on the initial conditions of the differential equation.  
Specifically, 
a differential equation is said to have the Painlev\'e property 
if all the movable singularities of all its solutions are poles.  
There is strong evidence~\cite{Yoshida83,Ziglin83a,Ziglin83b} that 
integrability is closely related to the singularity structure of the 
solutions of a differential equation (cf.~\cite{Ramani89,Steeb88}).  
For instance, dense branching of solutions around movable singularities 
has been shown to indicate nonintegrability~\cite{Ziglin82}.  

At the turn of the nineteenth-century, 
Painlev\'e~\cite{Painleve00} and his colleagues classified all the
rational second-order ODEs for which all the solutions are 
single-valued around all movable singularities. 
Equations possessing this property could either be solved in 
terms of known functions or transformed into one of the 
six Painlev\'e equations whose solutions define the Painlev\'e transcendents.
The Painlev\'e transcendents cannot be expressed in terms of the classical 
transcendental functions, 
except for special values of their parameters~\cite{Ince44}.

The complex singularity structure of solutions was first used by 
Kovalevskaya in 1889 to identify a new integrable 
system of equations for the motion for a rotating top 
(cf.~\cite{Grammaticos97,Steeb88}).  
Ninety years later, 
Ablowitz, Ramani and Segur (ARS)~\cite{Ablowitz80a,Ablowitz80b} 
and McLeod and Olver~\cite{McLeod83} 
formulated the Painlev\'e conjecture which gives a useful 
\emph{necessary} condition for determining whether a PDE is 
solvable using the Inverse Scattering Transform (IST) method.  
Specifically, the Painlev\'e conjecture asserts that every nonlinear ODE 
obtained by an exact reduction of a nonlinear PDE solvable by the IST-method 
has the Painlev\'e property.  
While necessary, the condition is not sufficient; 
in general, most PDEs do not have exact reductions to nonlinear ODEs and 
therefore satisfy the conjecture by default~\cite{Ward84}.
Weiss, Tabor and Carnevale (WTC)~\cite{Weiss83} proposed an algorithm for 
testing PDEs directly 
(which is analogous to the ARS algorithm for testing ODEs).  
% 
% @@@ begin change WH 01/03/2006 added reference to Goriely's book
% 
For a thorough discussion of the traditional Painlev\'e property, 
see \cite{Ablowitz91,Conte99,Ercolani91,Goriely01,Hone05,Newell87,% 
Pickering96,Ramani89,Steeb88,Tabor90}.
% 
% @@@ end change WH 01/03/2006

There are numerous methods for solving completely integrable nonlinear PDEs, 
for instance by explicit transformations into linear equations or 
by using the IST-method~\cite{Fokas87}.  
Recently, progress has been made using \emph{Mathematica} and \emph{Maple} 
in applying the IST-method to difficult equations, 
including the Camassa-Holm equation~\cite{Johnson03}.
While there is as yet no systematic way to determine if a differential 
equation is solvable using the IST-method~\cite{McLeod83},
having the Painlev\'e property is a strong indicator that it will be.

There are several implementations of the Painlev\'e test in various 
computer algebra systems, including \emph{Reduce}, \emph{Macsyma}, 
\emph{Maple} and \emph{Mathematica}.  
The implementations described in \cite{RandW86,Renner92,Scheen97} are limited 
to ODEs, while the implementations discussed in 
\cite{Hereman98,Xu03,Xu04,Xu05} 
allow the testing of PDEs directly using the WTC algorithm.  
The implementation for PDEs written in \emph{Mathematica} by 
Hereman et al.~\cite{Hereman98} is limited to two independent variables 
$(x$ and $t)$ and is unable to find all the dominant behaviors in systems 
with undetermined exponents $\alpha_i$ (as is the case with the 
Hirota-Satsuma system).  
Our package \texttt{PainleveTest.m}~\cite{painlevecode01} 
written in \emph{Mathematica} syntax, 
allows the testing of polynomial PDEs (and ODEs) 
with no limitation on the number of differential equations or 
the number of independent variables (except where limited by memory). 
Our implementation also allows the testing of differential equations 
that have undetermined dominant exponents $\alpha_i$ and 
that are parameterized by arbitrary functions (or constants).  
The implementations for PDEs written in \emph{Maple} by 
Xu and Li~\cite{Xu03,Xu04,Xu05} were written after 
the one presented in this paper and are comparable to our implementation. 

The paper is organized as follows: 
in Section~\ref{sec:analysis} we review the basics of Painlev\'e analysis.  
Section~\ref{sec:algorithm} discusses the WTC algorithm for testing PDEs 
and uses the Korteweg-de Vries (KdV) equation and 
the Hirota-Satsuma system of coupled KdV (cKdV) equations 
to show the subtleties of the algorithm.
We detail the algorithms to determine the dominant behavior, 
resonances, and constants of integration 
using a generalized system of coupled nonlinear Schr\"odinger (NLS) 
equations in Section~\ref{sec:keyAlgorithms}.  
Additional examples are presented in Section~\ref{sec:additionalExamples} 
to illustrate the capabilities of the software.  
Section~\ref{sec:compare} compares our software package to other codes 
and briefly discusses the generalizations of the WTC algorithm.  
The use of the package \texttt{PainleveTest.m}~\cite{painlevecode01} 
is shown in Section~\ref{sec:usage}.
We draw some conclusions and discuss the results in 
Section~\ref{sec:conclusion}.

%%%%%%%%%%%%%%%%%%%%%%%%%%%%%%%%%%%%%%%%%%%%%%%%%%%%%%%%%%%%%%%%
\section{Painlev\'e Analysis}
\label{sec:analysis}

Consider a system of $M$ polynomial differential equations, 
\begin{equation}
  \label{painleveSystem}
  F_i ( \vec{u}({\vec{z}}), \vec{u}^{\prime}({\vec{z}}), 
  \vec{u}^{\prime\prime}({\vec{z}}), \dotsc, 
  \vec{u}^{(m_i)} ({\vec{z}}) ) = \vec{0}, 
  \qquad i = 1,2,\dotsc,M,
\end{equation}
where the dependent variable $\vec{u}(\vec{z})$ has components 
$u_1(\vec{z}),\dotsc,u_M(\vec{z}),$ 
the independent variable $\vec{z}$ has components $z_1,\dotsc,z_N,$ and
$\vec{u}^{(k)}({\vec{z}})= \partial^k \vec{u}(\vec{z}) / %
(\partial z_1^{k_1} \partial z_2^{k_2} \dotsb \partial z_N^{k_N})$ 
denotes the collection of mixed derivative terms of order $k.$ 
Let $m = \sum_{i=1}^M m_i,$ % used for $r_m$ later, 
where $m_i$ is the highest order in each equation.
If there are any arbitrary coefficients (constants or analytic functions 
of $\vec{z})$ parameterizing the system, we assume they are nonzero.   
For simplicity, 
in the examples we denote the components of $\vec{u}(\vec{z})$ by 
$u(\vec{z}), v(\vec{z}), w(\vec{z}), \dotsc,$ and 
the components of $\vec{z}$ by $x,y,z,\dotsc,t.$

%\begin{definition}
A differential equation has the \emph{Painlev\'e property} if all the 
movable singularities of all its solutions are poles.  
%\end{definition}
%
A singularity is \emph{movable} if it depends on the constants of 
integration of the differential equation.  
For instance, the Riccati equation, 
\begin{equation}
  \label{riccati}
  w'(z) + w^2(z) = 0,
\end{equation}
has the general solution $w(z) = 1/(z-c),$ 
where $c$ is the constant of integration.  
Hence, (\ref{riccati}) has a movable simple pole at $z = c$ 
because it depends on the constant of integration.  
Solutions of ODEs can have various kinds of 
singularities, including branch points and essential singularities;  
examples of the various types of singularities~\cite{Kruskal97} are shown 
in Table~\ref{tbl:singularities}.
As a general property, 
solutions of \emph{linear} ODEs have only fixed singularities~\cite{Ince44}.

\begin{table}[htbp]
  \centering
  \begin{tabular}{c@{$\quad\Rightarrow\quad$}c} 
%    \multicolumn{2}{l}{Singularity Type} \\ 
%    Equation & General Solution \\ \hline 
    \multicolumn{2}{l}{Simple \emph{fixed} pole} \\ 
      $z\,w' + w = 0$ & $w(z) = c/z$ \\[1em] 
    \multicolumn{2}{l}{Simple \emph{movable} pole} \\
      $w' + w^2 = 0$ & $ w(z) = 1/(z - c)$ \\[1em]  
    \multicolumn{2}{l}{Movable \emph{algebraic branch} point} \\
      $2w w' - 1 = 0$ & $w(z) = \sqrt{z - c}$ \\[1em]  
    \multicolumn{2}{l}{Movable \emph{logarithmic branch} point} \\
      $w'' + {w'}^2 = 0$ & $w(z) = \log(z-c_1) + c_2$ \\[1em]  
    \multicolumn{2}{l}{Non-isolated movable \emph{essential} singularity} \\
      $(1+w^2)w'' + (1-2w){w'}^2 = 0$ & 
        $w(z) = \tan\{\ln(c_1 z + c_2)\}$ \\ 
  \end{tabular}
  \caption{Examples of various types of singularities.}
  \label{tbl:singularities}
\end{table}

In general, a function of several complex variables cannot have an 
isolated singularity~\cite{Osgood13}.  
For example, $f(z) = 1/z$ has an isolated singularity at the point $z=0,$ 
but the function $f(w,z) = 1/z$ of two complex variables, $w=u+iv, z=x+iy,$
has a two-dimensional manifold of singularities, 
namely the points $(u,v,0,0),$ 
in the four-dimensional space of these variables.  
Therefore, we will define a pole of a function of several complex 
variables as a point $(a_1,a_2,\dotsc,a_N),$ in whose neighborhood 
the function can be written in the form
$f(\vec{z}) = h(\vec{z})/g(\vec{z}),$
where $g$ and $h$ are both analytic in a region containing 
$(a_1,\dotsc,a_N)$ in its interior,
$g(a_1,\dotsc,a_N) = 0,$  and  $h(a_1,\dotsc,a_N) \neq 0.$

The WTC algorithm considers the singularity structure of the solutions 
around non-characteristic manifolds of the form
$ g(\vec{z}) = 0,$
where $g(\vec{z})$ is an analytic function of $\vec{z} = (z_1,z_2,\dotsc,z_N)$ 
in a neighborhood of the manifold.  
Specifically, if the singularity manifold is determined by $g(\vec{z}) = 0$ 
and $\vec{u}(\vec{z})$ is a solution of the PDE, 
then one assumes a Laurent series solution
\begin{equation}
  \label{laurent}
  u_i(\vec{z}) = g^{\alpha_i}(\vec{z}) 
    \sum_{k=0}^\infty u_{i,k}(\vec{z})g^k(\vec{z}), 
  \qquad i = 1,2,\dotsc,M,
\end{equation}
where the coefficients $u_{i,k}(\vec{z})$ are analytic functions of $\vec{z}$ 
with $u_{i,0}(\vec{z}) \not\equiv 0$ in a neighborhood of the 
manifold and the $\alpha_i$ are integers with at least 
one exponent $\alpha_i<0.$ 
The requirement that the manifold $g(\vec{z}) = 0$ is non-characteristic, 
ensures that the expansion (\ref{laurent}) is well defined in the 
sense of the Cauchy-Kovalevskaya theorem~\cite{Ward84,Weiss84}.  

Substituting (\ref{laurent}) into (\ref{painleveSystem}) and 
equating coefficients of like powers of $g(\vec{z})$ determines the 
possible values of $\alpha_i$ and defines a recursion relation 
for $u_{i,k}(\vec{z}).$  
The recursion relation is of the form
\begin{equation}
  \label{recursionRelation}
  Q_k \vec{u}_k 
    = \vec{G}_k(\vec{u}_0,\vec{u}_1,\dotsc,\vec{u}_{k-1},g,\vec{z}),
  \qquad \vec{u}_k = (u_{1,k},u_{2,k},\dotsc,u_{M,k})^T,
\end{equation}
where $Q_k$ is an $M\times M$ matrix and $T$ denotes transpose.

For (\ref{painleveSystem}) to pass the Painlev\'e test, 
the series (\ref{laurent}) should have $m - 1$ arbitrary 
functions as required by the Cauchy-Kovalevskaya theorem 
(as $g(\vec{z})$ is the $m$-th arbitrary function).  
If so, the Laurent series solution corresponds to the general solution of 
the equation~\cite{Ablowitz91}.
The $m-1$ arbitrary functions $u_{i,k}(\vec{z})$ occur when $k$ is 
one of the roots of $\det(Q_k).$   
These roots $r_1\le r_2\le \dotsb \le r_m$ are called \emph{resonances}.  
The resonances are also equal to the Fuchs indices of the 
auxiliary equations of Darboux~\cite{Conte93}.

Since the WTC algorithm is unable to detect essential singularities, 
it is only a necessary condition for the PDE to have the 
Painlev\'e property~\cite{Clarkson85}.  
While rarely done in practice, 
sufficiency is proved by finding a transformation which 
linearizes the differential equation, 
yields an auto-B\"acklund transformation, 
a B\"acklund transformation, or hodographic 
transformation~\cite{Hereman89JoPA} to another differential equation which 
has the Painlev\'e property 
(see \cite{Conte99,Kruskal97,Ramani89} for more information).  

%%%%%%%%%%%%%%%%%%%%%%%%%%%%%%%%%%%%%%%%%%%%%%%%%%%%%%%%%%%%%%%%
\section{Algorithm for the Painlev\'e Test}
\label{sec:algorithm}

% 
% @@@ begin change WH 01/03/2006 removed ODEs here, come back to ODEs later
% 
In this section, we outline the WTC algorithm for testing PDEs for the 
Painlev\'e property.  
We discuss the Kruskal simplification and the Painlev\'e test of ODEs after 
the three main steps are outlined.
% 
% @@@ end change WH 01/03/2006
% 
Each of these steps is illustrated using both the KdV equation and the 
cKdV equations due to Hirota and Satsuma.  
Details of the three main steps of the algorithm are postponed till 
Section~\ref{sec:keyAlgorithms}.

% % % % % % % % % % % % % % % % % % % % % % % % % % % % % % % % 
\step[Determine the dominant behavior]

It is sufficient to substitute
\begin{equation}
  \label{dominantBehavior}
  u_i(\vec{z}) = \chi_i g^{\alpha_i}(\vec{z}),
  \qquad i = 1,2,\dotsc,M,
\end{equation}
where $\chi_i$ is a constant, 
into (\ref{painleveSystem}) to determine the leading exponents $\alpha_i.$  
In the resulting polynomial system, equating every two or more possible lowest
exponents of $g(\vec{z})$ in each equation gives a linear system 
for $\alpha_i.$  
The linear system is then solved for $\alpha_i$ 
and each solution branch is investigated.  
The traditional Painlev\'e test requires that 
all the $\alpha_i$ are integers and that at least one is negative.

If any of the $\alpha_i$ are non-integer in a given branch, 
then that branch of the algorithm terminates.  
A non-integer $\alpha_i$ implies that some solutions 
of (\ref{painleveSystem}) have movable algebraic branch points.  
Often, a suitable change of variables in (\ref{painleveSystem}) can 
remove the algebraic branch point.  
An alternative approach is to use the ``weak'' Painlev\'e test, 
which allows certain rational $\alpha_i$ and resonances;  
% 
% @@@ begin change WH 01/03/2006 added reference to Goriely's book
% 
see~\cite{Goriely01,Hone05,Ramani82,Ramani89} for more information.
% on the weak Painlev\'e test.
% 
% @@@ end change WH 01/03/2006
% 

If one or more $\alpha_i$ remain undetermined, we assign 
integer values to the free $\alpha_i$ 
so that every equation in (\ref{painleveSystem}) has at least 
two different terms with equal lowest exponents.

For each solution $\alpha_i,$ we substitute 
\begin{equation}
  \label{dominantBehavior2}
  u_i(\vec{z}) = u_{i,0}(\vec{z}) g^{\alpha_i}(\vec{z}),
  \qquad i = 1,2,\dotsc,M,
\end{equation}
into (\ref{painleveSystem}).  
We then solve the (typically) nonlinear equation for $u_{i,0}(\vec{z}),$ 
which is found by balancing the leading terms.  
By leading terms, we mean those terms with the lowest exponent of 
$g(\vec{z}).$ 
If any of the solutions contradict the assumption that 
$u_{i,0}(\vec{z}) \not\equiv 0,$ 
then that branch of the algorithm fails the Painlev\'e test.

If any of the $\alpha_i$ are non-integer,  
all the $\alpha_i$ are positive, or there is a contradiction with the 
assumption that $u_{i,0}(\vec{z}) \not\equiv 0,$
then that branch of the algorithm terminates 
and does not pass the Painlev\'e test for that branch.  

% % % % % % % % % % % % % % % % % % % % % % % % % % % % % % % % 
\step[Determine the resonances]

For each $\alpha_i$ and $u_{i,0}(\vec{z})$, we calculate the 
$r_1\le \dotsb \le r_m$ for which $u_{i,r}(\vec{z})$ is an 
arbitrary function in (\ref{laurent}).  
To do this, we substitute
\begin{equation}
  \label{resonancesAnsatz}
  u_i(\vec{z}) = u_{i,0}(\vec{z}) g^{\alpha_i}(\vec{z}) + u_{i,r}(\vec{z}) 
  g^{\alpha_i + r}(\vec{z}) 
\end{equation}
into (\ref{painleveSystem}), 
and keep only the lowest order terms in $g(\vec{z})$ 
that are linear in $u_{i,r}.$ 
This is done by computing the solutions for $r$ of $\det(Q_r) = 0,$ 
where the $M \times M$ matrix $Q_r$ satisfies
\begin{equation}
  \label{qMatrix}
  Q_r \vec{u}_r = \vec{0}, \qquad 
  \vec{u}_r = (u_{1,r} \,\; u_{2,r} \,\; \dotsc \,\; u_{M,r})^T.
\end{equation}

If any of the resonances are non-integer, 
then the Laurent series solutions of (\ref{painleveSystem}) have a movable 
algebraic branch point and the algorithm terminates.  
If $r_m$ is not a positive integer, then the algorithm terminates; 
if $r_1 = -1, r_2 = \dotsb = r_m = 0$ and $m-1$ of the $u_{i,0}(\vec{z})$ 
found in Step 1 are arbitrary, 
then (\ref{painleveSystem}) passes the Painlev\'e test.  
If (\ref{painleveSystem}) is parameterized, 
the values for $r_1\le \dotsb \le r_m$ may depend on the parameters,
and hence restrict the allowable values for the parameters.  

There is always a resonance $r=-1$ which corresponds to 
the arbitrariness of $g(\vec{z});$  
as such, it is often called the universal resonance.  
When there are negative resonances other than $r=-1,$ 
(or, more than one resonance equals $-1)$
then the Laurent series solution is not the general solution and 
further analysis is needed to determine if (\ref{painleveSystem}) 
passes the Painlev\'e test.    
The perturbative Painlev\'e approach,  
developed by Conte et al.~\cite{CFP93},
is one method for investigating negative resonances.  

% % % % % % % % % % % % % % % % % % % % % % % % % % % % % % % % 
\step[Find the constants of integration and check compatibility conditions]

For the system to possess the Painlev\'e property, the arbitrariness of 
$u_{i,r}(\vec{z})$ must be verified up to the highest resonance level.   
This is done by substituting
\begin{equation}
  \label{constantsOfIntegration}
  u_i(\vec{z}) = g^{\alpha_i}(\vec{z}) 
    \sum_{k = 0}^{r_m} u_{i,k}(\vec{z}) g^k(\vec{z})
\end{equation}
into (\ref{painleveSystem}), 
where $r_m$ is the largest positive integer resonance.  

For (\ref{painleveSystem}) to have the Painlev\'e property, 
the $(M+1) \times M$ augmented matrix $(Q_k|\vec{G}_k)$
must have rank $M$ when $k \neq r$ and rank $M-s$ when $k=r,$ 
where $s$ is the algebraic multiplicity of $r$ in $\det(Q_r)=0,$ 
$1 \le k \le r_m,$ and $Q_k$ and $\vec{G}_k$ are as 
defined in (\ref{recursionRelation}).  
If the augmented matrix $(Q_k|\vec{G}_k)$ has the correct rank, 
solve the linear system (\ref{recursionRelation}) for 
$u_{1,k}(\vec{z}),\dotsc,u_{M,k}(\vec{z})$ and use the results in 
the linear system at level $k+1.$  

If the linear system (\ref{recursionRelation}) does not have a solution, 
then the Laurent series solution of (\ref{painleveSystem}) has a 
movable logarithmic branch point and the algorithm terminates.  
Often, when (\ref{painleveSystem}) is parameterized, 
carefully choosing the parameters will resolve the difference in 
the ranks of $Q_k$ and $(Q_k|\vec{G}_k).$

If the algorithm does not terminate, 
then the Laurent series solutions of (\ref{painleveSystem}) are free 
of movable algebraic or logarithmic branch points 
and (\ref{painleveSystem}) passes the Painlev\'e test.  

% 
% @@@ begin change WH 01/03/2006 added `of PDEs' 
% 
The Painlev\'e test of PDEs is quite cumbersome; in particular, Step 3 is 
lengthy and prone to error when done by hand.  
% 
% @@@ begin change WH 01/03/2006 bares ---> bears
% 
To simplify Step 3, Kruskal proposed a simplification which now bears his 
name.  
% 
% @@@ end change WH 01/03/2006 
% 
In the context of the WTC algorithm, it is sometimes called the 
Weiss-Kruskal simplification~\cite{Jimbo82,Kruskal97}.
% 
% The Weiss-Kruskal simplification~\cite{Jimbo82,Kruskal97}
% greatly simplifies the computation of the constants of integration.  
%
The manifold defined by $g(\vec{z}) = 0$ is non-characteristic, 
that means $g_{z_l}(\vec{z}) \neq 0$ for some $l$ on the manifold 
$g(\vec{z}) = 0.$  
By the implicit function theorem, we can then locally solve $g(\vec{z}) = 0$ 
for $z_l,$ so that 
\begin{equation}
  \label{krusalsimp}
  g(\vec{z}) = z_l - h(z_1,\dotsc,z_{l-1},z_{l+1},\dotsc,z_N),
\end{equation}
for some arbitrary function $h.$  
Using (\ref{krusalsimp}) greatly simplifies the computation of 
% 
% @@@ begin change WH 01/03/2006 arguments of integration constants 
% 
the constants of integration $u_{i,k}(z_1,\dotsc,z_{l-1},z_{l+1},\dotsc,z_N).$
% 
%  which are now independent of $z_l.$  
% 
% @@@ end change WH 01/03/2006 
% 
% @@@ begin change WH 01/03/2006 reformulated
% 
However, with the Kruskal simplification one looses the ability to use the 
Weiss truncation method~\cite{Weiss83b} to find a linearising transformation, 
an auto-B\"acklund transformation, or a B\"acklund transformation 
(see \cite{Conte99}).
% 
% @@@ end change WH 01/03/2006
% 

% 
% @@@ begin change WH 01/03/2006 added ODE issues here
% @@@ begin change DB 1/4/2006, changed g to g(z).
%
When testing ODEs, (\ref{laurent}) must be replaced by 
\begin{equation}
  \label{laurentODE}
  u_i(z) = g^{\alpha_i}(z) 
    \sum_{k=0}^\infty u_{i,k} \, g^k(z), 
  \qquad i = 1,2,\dotsc,M,
\end{equation}
where the coefficients $u_{i,k}$ are constants, $g(z) = z - z_0,$ 
and $z_0$ is an arbitrary constant.
If $z$ explicitly occurs in the ODE, then it is (automatically) replaced by 
$g(z) + z_0$ prior to Step 1 of the test.
An example of the Painlev\'e test of an ODE is given in 
Section~\ref{sec:additionalExamples}.
%
% @@@ end change DB 1/4/2006
% @@@ end change WH 01/03/2006

% % % % % % % % % % % % % % % % % % % % % % % % % % % % % % % % 
\subsection{The Korteweg-de Vries equation}
\label{sec:kdv}

To illustrate the steps of the algorithm, 
let us examine the KdV equation~\cite{Ablowitz91}, 
\begin{equation}
  \label{kdv}
  u_t + 6 uu_x + u_{3x} = 0,
\end{equation}
the most famous completely integrable PDE from soliton theory.  
Note that for simplicity, 
we use $u_{ix} = u_{xx\dotsb x} = \partial^i u/\partial x^i$ 
and $g_{ix} = \partial^i g/\partial x^i$
when $i \ge 3.$

Substituting (\ref{dominantBehavior}) into (\ref{kdv}) gives
\begin{equation}
  \alpha \chi \left\{ 
    g_t g^{\alpha - 1} \!+\! 6 \chi g_x g^{2\alpha - 1} \!+\! 
      g^{\alpha - 3} [ (\alpha\!-\!1)\left((\alpha\!-\!2) g_x^2 
        \!+\! 3 gg_{xx}\right)g_x \!+\! g^2g_{3x}]
  \right\} = 0.
\end{equation}
The lowest exponents of $g(x,t)$ are $\alpha-3$ and $2\alpha-1.$  
Equating these leading exponents gives $\alpha = -2.$  
%Thus, there is a double pole at $g(\vec{z}) = 0$.
Substituting (\ref{dominantBehavior2}), $u(x,t) = u_0(x,t) g^{-2}(x,t),$ 
into (\ref{kdv}) and requiring that the leading terms (in $g^{-5}(x,t))$ 
balance, gives $u_0(x,t) = -2g_x^2(x,t).$  

Substituting (\ref{resonancesAnsatz}), 
$u(x,t) = -2g_x^2(x,t) g^{-2}(x,t) + u_r(x,t) g^{r-2}(x,t),$ 
into (\ref{kdv}) and 
equating the coefficients of the dominant terms (in $g^{r-5}(x,t))$ 
that are linear in $u_r(x,t)$ gives 
\begin{equation}
  (r-6)(r-4)(r+1)g_x(x,t)^3 = 0.
\end{equation}
Assuming $g_x(x,t) \not\equiv 0,$ 
the resonances of (\ref{kdv}) are $r_1=-1,r_2 = 4$ and $r_3 = 6.$  

We now substitute 
\begin{gather} \notag
  u(x,t) = g^{-2}(x,t) \sum_{k=0}^6 u_{k}(x,t)g^k(x,t) \\ 
    \qquad = -2g_x^2(x,t) g^{-2}(x,t) + u_1(x,t) g^{-1}(x,t) 
    + \dotsb + u_6(x,t) g^{4}(x,t)
\end{gather}
into (\ref{kdv}) and group the terms of like powers of $g(x,t).$  
So, we will pull off the coefficients of $g^{k-5}(x,t)$ at level $k.$ 
Equating the coefficients of $g^{-4}(x,t)$ to zero at level $k \!=\! 1,$ 
gives $u_1(x,t) g_x^3(x,t) \!=\! 2 g_x^3(x,t)g_{xx}(x,t).$ 
% for $u_1(x,t).$
Setting $u_1(x,t) \!=\! 2g_{xx}(x,t),$ we get 
\begin{equation}
  u_2(x,t) = -\frac{ g_t g_x^2 + 3 g_x g_{xx}^2 - 4g_x^2g_{3x}}{6g_x^3},
\end{equation}
at level $k \!=\! 2.$ 
Similarly, at level $k \!=\! 3,$ 
\begin{equation}
  u_3(x,t) = \frac{g_x^2 g_{xt} - g_t g_x g_{xx} + 3g_{xx}^3 - 
4g_xg_{xx}g_{3x}+g_x^2g_{4x}}{6 g_x^4}.
\end{equation}
At level $k = r_2 = 4,$ we find
\begin{equation}
  \label{kdvr2}
  (u_1)_t + 6\{u_3(u_0)_x + u_2(u_1)_x + u_1(u_2)_x + u_0(u_3)_x\} 
    + (u_1)_{3x} = 0,
\end{equation}
which is trivially satisfied upon substitution of 
the solutions of $u_0(x,t),\dotsc,u_3(x,t).$
Therefore, the compatibility condition at level $k=r_2=4$ is satisfied 
and $u_4(x,t)$ is indeed arbitrary.  
At level $k=5,$ $u_5(x,t)$ is unambiguously determined, 
but not shown due to length.  
Finally, the compatibility condition at level $k=r_3=6$ is trivially satisfied 
when the solutions for $u_0(x,t), \dotsc, u_3(x,t)$ and $u_5(x,t)$ are 
substituted into the recursion relation at that resonance level.  

Therefore, 
the Laurent series solution $u(x,t)$ of (\ref{kdv}) in 
the neighborhood of $g(x,t) = 0$ is free of algebraic 
and logarithmic movable branch points.  
Furthermore, since the Laurent series solution,
\begin{equation}
  u(x,t) = g^{-2}(x,t) \sum_{k=0}^\infty u_k(x,t) g^k(x,t),
\end{equation}
has three arbitrary functions, $g(x,t),u_4(x,t),$ and $u_6(x,t),$ 
(as required by the Cauchy-Kovalevskaya theorem 
since (\ref{kdv}) is of third order) it is also the general solution.  
Hence, we conclude that (\ref{kdv}) passes the Painlev\'e test.

% 
% @@@ begin change WH 01/03/2006 added simplified series and 
%     also replaced u_4(x,t), u_6(x,t) by u_4(t), u_6(t).
% 
The Weiss-Kruskal simplification uses $g(x,t) = x - h(t).$ 
Consequently, $g_x = 1, g_{xx} = g_{3x} = \cdots = 0,$ 
and the Laurent series,  
\begin{equation}
  \label{laurentKruskla}
  u(x,t) = g^{-2}(x,t) 
    \sum_{k=0}^\infty u_{i,k}(t) \, g^k(x,t), 
\end{equation}
becomes 
\begin{multline}
  u(x,t) = -\frac{2}{(x-h(t))^2} + \frac{1}{6}h'(t) + u_4(t) (x-h(t))^2 \\
    + \frac{1}{36} h''(t) (x-h(t))^3 + u_6(t) (x-h(t))^4 + \dotsb,
\end{multline}
where $h(t), u_4(t)$ and $u_6(t)$ are arbitrary.  
% 
% @@@ end change WH 01/03/2006
% 

% % % % % % % % % % % % % % % % % % % % % % % % % % % % % % % % 
\subsection{The Hirota-Satsuma system}

To show the subtleties in determining the dominant behavior,  
consider the cKdV equations due to Hirota and Satsuma~\cite{Ablowitz91} 
with real parameter $a,$
\begin{equation}
  \label{hirota}
  \begin{aligned}
    u_t & = a (6 u u_x + u_{3x}) - 2 v v_x, && a > 0, \\
    v_t & = - 3 u v_x - v_{3x}.
  \end{aligned} 
\end{equation} 
%which model shallow water waves.
Again, we substitute (\ref{dominantBehavior}), 
$u(x,t) = \chi_1 g^{\alpha_1}(x,t)$ and $v(x,t) = \chi_2 g^{\alpha_2}(x,t),$
into (\ref{hirota}) and pull off the lowest exponents of $g(x,t).$ 
>From the first equation, 
we get $\alpha_1 - 3,  2\alpha_1 - 1,$ and $2\alpha_2 - 1.$ 
>From the second equation, 
we get $\alpha_2 - 3$ and  $\alpha_1 + \alpha_2 - 1.$
Hence, $\alpha_1 = -2$ from the second equation.  
Substituting this into the first equation gives $\alpha_2 \ge -2.$  

Substituting (\ref{dominantBehavior2}) into (\ref{hirota}) and 
requiring that at least two  leading terms balance gives us two branches:
$\alpha_1 = \alpha_2 = -2$ and $\alpha_1 = -2,\alpha_2 = -1.$
The branches with $\alpha_1 = -2$ and $\alpha_2 \ge 0$ 
are excluded for they require that either $u_0(x,t)$ or $v_0(x,t)$ 
is identically zero.  

Continuing with the two branches and 
solving for $u_0(x,t)$ and $v_0(x,t)$ gives 
\begin{equation}
  \begin{cases}
    \alpha_1 = \alpha_2 = -2, \\   
    u_0(x,t) = -4 g_x^2(x,t), \\
    v_0(x,t) = \pm 2 \sqrt{6a}g_x^2(x,t),
  \end{cases} 
  \quad \text{and} \qquad 
  \begin{cases}
    \alpha_1 = -2,~\alpha_2 = -1, \\
    u_0(x,t) = -2g_x^2(x,t), \\
    v_0(x,t) \text{ arbitrary}.
  \end{cases}
\end{equation}

For the branch with $\alpha_1=\alpha_2=-2,$ 
substituting (\ref{resonancesAnsatz}),
\begin{equation}
  \begin{cases}
    u(x,t) = -4 g_x^2(x,t) g^{-2}(x,t) + u_r(x,t) g^{r-2}(x,t), \\
    v(x,t) = \pm 2 \sqrt{6a}g_x^2(x,t) g^{-2}(x,t) + v_r(x,t) g^{r-2}(x,t),
  \end{cases}
\end{equation}
into (\ref{hirota}) and 
equating to zero the coefficients of the lowest order terms in $g(x,t)$ 
that are linear in $u_r$ and $v_r$ gives 
\begin{equation}
  \begin{pmatrix}
    -(r - 4)(r^2 - 5r - 18)a g_x^3(x,t) 
      & \pm 12 \sqrt{6 a} g_x^3(x,t) \\
    \mp 4 (r - 4) \sqrt{6 a}g_x^3(x,t) 
      & (r - 2)(r - 7) r g_x^3(x,t)
  \end{pmatrix}
  \begin{pmatrix}
    u_r(x,t) \\ v_r(x,t)
  \end{pmatrix}
  = \vec{0}.
\end{equation}
>From 
\begin{equation}
  \det(Q_r)  = -a (r + 2)(r + 1)(r - 3)(r - 4)(r - 6)(r - 8) g_x^6(x,t) = 0,
\end{equation}
we obtain the resonances $r_1=-2, r_2=-1, r_3=3, r_4=4, r_5=6,$ and $r_6=8.$  

By convention, 
the resonance $r_1=-2$ is ignored since it violates the hypothesis that 
$g(x,t)^{-2}$ is the dominant term in the expansion near $g(\vec{z}) = 0$.
Furthermore, this is not a principal branch since the series has 
only five arbitrary functions instead of the required six 
(as the term corresponding to resonance $r_1=-2$ does not contribute 
to the expansion).  
Thus, this leads to a particular solution 
and the general solution may still be multi-valued.  
%To investigate negative resonances, Conte et al~\cite{CFP93}
%developed the perturbative Painlev\'e approach,
%which is beyond the scope of this paper.

As in the previous example, 
the constants of integration at level $k$ are found by substituting 
(\ref{constantsOfIntegration}) into (\ref{hirota}) and pulling off 
the coefficients of $g^{k-5}(x,t).$  
At level $k= 1,$ 
\begin{equation}
  \begin{pmatrix}
      11 a g_x^3(x,t) & \pm 2\sqrt{6a} g_x^3(x,t) \\
      \pm 2 \sqrt{6a}g_x^3(x,t) & -g_x^3(x,t) 
  \end{pmatrix}
  \begin{pmatrix}
      u_1(x,t) \\
      v_1(x,t)
  \end{pmatrix}
  =
  \begin{pmatrix}
      20 a g_x^3(x,t) g_{xx}(x,t) \\
     \pm 10 \sqrt{6a}g_x^3(x,t)g_{xx}(x,t)
  \end{pmatrix},
\end{equation}
and thus,
\begin{equation}
  u_1(x,t) = 4 g_{xx}(x,t), \qquad 
    v_1(x,t) = \pm 2 \sqrt{6a} g_{xx}(x,t).
\end{equation}
At level $k= 2$,
\begin{equation}
  \begin{aligned}
  u_2(x,t) & = \frac{3 g_{xx}^2(x,t) - g_x(x,t)\big(g_t(x,t) 
  + 4 g_{3x}(x,t)\big)}{3 g_x^2(x,t)}, \\
    v_2(x,t) & = 
      \pm \frac{(1 + 2a) g_t(x,t) g_x(x,t) + 4 a g_x(x,t)g_{3x}(x,t) 
        -3 a g_{xx}^2(x,t) }{\sqrt{6 a } g_x^2(x,t)}.
  \end{aligned} 
\end{equation}
The compatibility conditions at levels $k = r_3 = 3$ and $k = r_4 = 4$ are 
trivially satisfied.  
At levels $k=5$ and $k=7,$ 
$u_k(x,t)$ and $v_k(x,t)$ are unambiguously determined (not shown). 
At resonance levels $k=r_5=6$ and $k=r_6=8,$ 
the compatibility conditions require $a = \frac{1}{2}.$ \\

Likewise, for the branch with $\alpha_1=-2,\alpha_2=-1,$  
substituting (\ref{resonancesAnsatz}),
\begin{equation}
  \begin{cases}
    u(x,t) = -2 g_x^2(x,t) g^{-2}(x,t) + u_r(x,t) g^{r-2}(x,t), \\
    v(x,t) = v_0(x,t) g^{-1}(x,t) + v_r(x,t) g^{r-1}(x,t),
  \end{cases}
\end{equation}
into (\ref{hirota}) gives
\begin{equation}
  \begin{pmatrix}
    -a(r+1)(r-4)(r-6)g_x^3(x,t) & -3v_0(x,t) g_x(x,t) \\
    0 & r(r-1)(r-5)g_x^3(x,t)
  \end{pmatrix}
  \begin{pmatrix}
    u_r(x,t) \\ v_r(x,t)
  \end{pmatrix}
  = \vec{0}.
\end{equation}
Since 
\begin{equation}
  \det(Q_r)  = -a (r+1)r(r-1)(r-4)(r-5)(r-6) g_x^6(x,t),
\end{equation}
the resonances are $r_1=-1, r_2=0, r_3=1, r_4=4, r_5=5,$ and $r_6=6.$  

Since $r_2=0$ is a resonance, 
there must be freedom at level $k=r_2=0;$
indeed, 
coefficient $u_0(x,t) = -2 g_x^2(x,t)$ is unambiguously determined 
but $v_0(x,t)$ is arbitrary.  
Then, the constants of integration are found by substituting 
%(\ref{constantsOfIntegration}), i.e. 
\begin{equation}
  \begin{cases}
    u(x,t) = -2 g_x^2(x,t) g^{-2}(x,t) + u_1(x,t) g^{-1}(x,t) 
      + \dotsb + u_6(x,t)g^4(x,t), \\
    v(x,t) = v_0(x,t) g^{-1}(x,t) + v_1(x,t) + \dotsb + v_6(x,t) g^5(x,t),
  \end{cases}
\end{equation}
into (\ref{hirota}) and pulling off the coefficients of 
$g^{k-5}(x,t)$ in the first equation and $g^{k-4}(x,t)$ in the second equation.
At level $k = r_3 = 1,$ 
\begin{equation}
  \begin{pmatrix}
    a  & 0 \\
    v_0(x,t) & 0 
  \end{pmatrix}
  \begin{pmatrix}
    u_1(x,t) \\
    v_1(x,t)
  \end{pmatrix}
  =   
  \begin{pmatrix}
    2a g_{xx}(x,t) \\
    2 v_0(x,t) g_{xx}(x,t)
  \end{pmatrix}. 
\end{equation}
So, $u_1(x,t) = 2g_{xx}(x,t)$ and $v_1(x,t)$ is arbitrary.  
At level $k= 2,$
\begin{equation}
  \begin{pmatrix}
    12 a g_x^2 & 0 \\
    -3v_0  g_x  & -6g_x ^3
  \end{pmatrix}
  \begin{pmatrix}
      u_2  \\
      v_2 
  \end{pmatrix}
  = 
  \begin{pmatrix}
    2g_t g_x \!+\! 6 a g_{xx} ^2 \!-\! v_0 ^2
      \!-\! 8ag_x g_{3x}  \\
    v_0 g_t  \!+\! 6 (v_1)_xg_x ^2   \!-\! 3 (v_0)_x g_{xx}  
      \!+\! 3(v_0)_{2x}  g_x  \!+\! v_0  g_{3x} 
  \end{pmatrix},
\end{equation}
which unambiguously determines $u_2(x,t)$ and $v_2(x,t).$  
Similarly, the coefficients in the Laurent series solution are 
unambiguously determined at level $k=3.$  
At resonance level $k = r_4 = 4,$ the compatibility condition is 
trivially satisfied.  
At resonance levels $k= r_5 = 5$ and $k = r_6 = 6,$ 
the compatibility conditions requires $a = \frac{1}{2}.$  

Therefore, (\ref{hirota}) satisfies the necessary conditions for having 
the Painlev\'e property when $a = \frac{1}{2},$
a fact confirmed by other analyses of complete integrability~\cite{Ablowitz91}.

%%%%%%%%%%%%%%%%%%%%%%%%%%%%%%%%%%%%%%%%%%%%%%%%%%%%%%%%%%%%%%%%
\section{Key Algorithms}
\label{sec:keyAlgorithms}

In this section,
we present the three key algorithms in greater detail.  
To illustrate the steps we consider a generalization 
\begin{equation} 
  \label{coupledNLS}
  \begin{gathered}
  iu_t + u_{xx} + (|u|^2 + \beta |v|^2)u + a(x,t) u + c(x,t) v = 0, \\
  iv_t + v_{xx} + (|v|^2 + \beta |u|^2)v + b(x,t) v + d(x,t) u = 0,
  \end{gathered}
\end{equation}
of the coupled nonlinear Schr\"odinger (NLS) equations~\cite{Tan01}
\begin{equation} 
  \begin{gathered}
  iu_t + u_{xx} + (|u|^2 + \beta |v|^2)u = 0, \\
  iv_t + v_{xx} + (|v|^2 + \beta |u|^2)v = 0.
  \end{gathered}
\end{equation}
In (\ref{coupledNLS}), 
$a(x,t),\dotsc,d(x,t)$ are arbitrary complex functions 
and $\beta$ is a real constant parameter.  
Since all the functions in (\ref{coupledNLS}) are complex, 
we write the system as
\begin{equation} 
  \label{genNLS}
  \begin{gathered}
  iu_t + u_{xx} + (u\bar{u} + \beta v\bar{v})u + a(x,t) u + c(x,t) v = 0, \\
  i\bar{u}_t - \bar{u}_{xx} - 
  (u\bar{u} + \beta v\bar{v})\bar{u} - \bar{a}(x,t) \bar{u} - 
  \bar{c}(x,t)\bar{v} = 0, \\
  iv_t + v_{xx} + (v\bar{v} + \beta u\bar{u})v + b(x,t) v + d(x,t) u = 0, \\
  i\bar{v}_t - \bar{v}_{xx} - 
  (v\bar{v} + \beta u\bar{u})\bar{v} - \bar{b}(x,t) \bar{v} - 
  \bar{d}(x,t) \bar{u} = 0,
  \end{gathered}
\end{equation}
and treat $u,\bar{u},v,$ and $\bar{v}$ as independent complex functions.  
As is customary, the variables with overbars denote complex conjugates.  
%$\bar{a}(x,t),\dotsc,\bar{d}(x,t)$ are the complex conjugates 
%of $a(x,t),\dotsc,d(x,t).$

% % % % % % % % % % % % % % % % % % % % % % % % % % % % % % % % 
\subsection{Algorithm to determine the dominant behavior}
\label{algo:dominantBehavior}

Determining the dominant behavior of (\ref{painleveSystem}) is 
delicate and the omission of valid dominant behaviors often leads 
to erroneous results~\cite{Ramani89}.

%  %  %  %  %  %  %  %  %  %  %  %  %  %  %  %  %  %  %  %  %  %
\step[Substitute the leading-order ansatz]

To determine the values of $\alpha_i,$ 
it is sufficient to substitute
$  u_i(\vec{z}) = \chi_i g(\vec{z})^{\alpha_i}, $
into (\ref{painleveSystem}), where $\chi_i$ is constant 
and $g(\vec{z})$ is an analytic function in a neighborhood 
of the non-characteristic manifold defined by $g(\vec{z}) = 0.$  

%  %  %  %  %  %  %  %  %  %  %  %  %  %  %  %  %  %  %  %  %  %
\step[Collect exponents and prune non-dominant branches]

The balance of exponents must come from different terms in 
(\ref{painleveSystem}).  
For each equation $F_i=0,$ collect the exponents of $g(\vec{z}).$  
Then, remove non-dominant exponents and duplicates 
(that come from the same term in (\ref{painleveSystem})). 
For example, $\alpha_1 + 1$ is non-dominant and can be removed from 
$\{ \alpha_1 - 1, \alpha_1 + 1\}$ since $\alpha_1 -1 < \alpha_1 + 1.$  

For (\ref{genNLS}), the exponents corresponding to each equation are
\begin{equation}
  \label{genNLSalphaSystem}
  \begin{aligned}
    & F_1: \quad 
    \{ \alpha_1-2, 2\alpha_1 + \alpha_2, \alpha_1 + \alpha_3 + \alpha_4 \}, \\
    & F_2: \quad 
    \{ \alpha_2-2, \alpha_1 + 2\alpha_2, \alpha_2 + \alpha_3 + \alpha_4 \}, \\
    & F_3: \quad 
    \{ \alpha_3-2, 2\alpha_3 + \alpha_4, \alpha_1 + \alpha_2 + \alpha_3 \}, \\
    & F_4: \quad 
    \{ \alpha_4-2, \alpha_3 + 2\alpha_4, \alpha_1 + \alpha_2 + \alpha_4 \},
  \end{aligned}
\end{equation}
after duplicates and non-dominant exponents have been removed.

%  %  %  %  %  %  %  %  %  %  %  %  %  %  %  %  %  %  %  %  %  %
\step[Combine expressions and compute relations for $\alpha_i$]

For each $F_i$ separately, equate all possible combinations of two 
elements.  
Then, construct relations between the $\alpha_i$ by solving for 
$\alpha_1,\alpha_2,$ etc., one at a time.

For (\ref{genNLSalphaSystem}), we get
\begin{equation}
  \begin{gathered}
    F_1: \quad 
    \{ \alpha_1-2 = 2\alpha_1 + \alpha_2, 
    \alpha_1-2 = \alpha_1 + \alpha_3 + \alpha_4, \\
    2\alpha_1 + \alpha_2 = \alpha_1 + \alpha_3 + \alpha_4 \} \\
    \qquad 
    \Rightarrow \{ \alpha_1 + \alpha_2 = -2, \alpha_3+\alpha_4 = -2, 
	\alpha_1 + \alpha_2 = \alpha_3 + \alpha_4 \}.
  \end{gathered}
\end{equation}
For $F_2,F_3$ and $F_4$ we again find that 
$\{ \alpha_1 + \alpha_2 = -2, \alpha_3+\alpha_4 = -2, 
\alpha_1 + \alpha_2 = \alpha_3 + \alpha_4 \}.$  

%  %  %  %  %  %  %  %  %  %  %  %  %  %  %  %  %  %  %  %  %  %
\step[Combine equations and solve for exponents $\alpha_i$]

By combining the sets of expressions in an ``outer product'' fashion,
we generate all the possible linear equations for $\alpha_i.$  
Solving these linear systems, we form a set of all possible solutions 
for $\alpha_i.$  

For (\ref{genNLS}), we have three sets of linear equations
\begin{gather}
  \begin{cases}
    \alpha_1 + \alpha_2 = -2
  \end{cases} \Rightarrow 
  \begin{cases}
    \alpha_1 + \alpha_2 = -2, \\
    \alpha_3 + \alpha_4 \ge -2,
  \end{cases} \\
  \begin{cases}
    \alpha_3 + \alpha_4 = -2
  \end{cases} \Rightarrow 
  \begin{cases}
    \alpha_1 + \alpha_2 \ge -2, \\
    \alpha_3 + \alpha_4 = -2,
  \end{cases} 
\intertext{and}
  \begin{cases}
    \alpha_1 + \alpha_2 = \alpha_3 + \alpha_4,
  \end{cases} \Rightarrow 
  \begin{cases}
    \alpha_1 + \alpha_2 = \alpha_3 + \alpha_4 \ge -2. \\
  \end{cases} 
\end{gather}

Although the algorithm treats $u,\bar{u},v,$ and $\bar{v}$ 
as independent complex functions, 
we know that $\alpha_1 = \alpha_2$ and $\alpha_3 = \alpha_4$ 
because $\bar{u}$ and $\bar{v}$ are the complex conjugates 
of $u$ and $v.$  
Our package \texttt{PainleveTest.m} can take advantage 
of such additional information by using the option 
\texttt{DominantBehaviorConstraints -> 
\{alpha[1] == alpha[2], alpha[3] == alpha[4]\}.}
Using this additional information yields three cases, 
$\alpha_1 = \alpha_2 = \alpha_3 = \alpha_4 \ge -1$ and
$\alpha_1 = \alpha_2 = -1, \alpha_3 = \alpha_4 \ge -1$ and
$\alpha_1 = \alpha_2 \ge -1, \alpha_3 = \alpha_4 = -1.$

%  %  %  %  %  %  %  %  %  %  %  %  %  %  %  %  %  %  %  %  %  %
\step[Fix the undetermined $\alpha_i$]

First, compute the minimum values for the undetermined $\alpha_i.$ 
If a minimum value cannot be determined, then the user-defined 
value \texttt{DominantBehaviorMin} is used.  
If so, the value of the free $\alpha_i$ is counted up to a user defined 
\texttt{DominantBehaviorMax}.  
% 
% @@@ begin change WH 01/03/2006 added a line about DominantBehavior
% 
If neither of the bounds is set, the software will run the test for
the default values $\alpha_i = -1, -2$ and $-3.$
For maximal flexibility, with the option \texttt{DominantBehavior} one can 
also run the code for user-specified values of $\alpha_i.$ 
An example is given in Section~\ref{sec:usage}.  
% 
% @@@ end change WH 01/03/2006
% 
% @@@ begin change WH 01/03/2006 reformulated
% 
In any case, the selected or given dominant behaviors are checked for 
consistency with (\ref{painleveSystem}).  
% 
% @@@ end change WH 01/03/2006 
% 

For (\ref{genNLS}), if we take $\alpha_1,\dotsc,\alpha_4 < 0,$ 
then we are left with only one branch 
\begin{equation}
    \alpha_1 = \alpha_2 = \alpha_3 = \alpha_4 = -1.
\end{equation}

%  %  %  %  %  %  %  %  %  %  %  %  %  %  %  %  %  %  %  %  %  %
\step[Compute the first terms in the Laurent series]

Using the values for $\alpha_i,$ substitute 
\begin{equation}
  u_i(\vec{z}) = u_{i,0}(\vec{z}) g^{\alpha_i}(\vec{z})
\end{equation}
into (\ref{painleveSystem}) 
and solve the resulting (typically) nonlinear equations for 
$u_{i,0}(\vec{z})$ using the assumption that $u_{i,0}(\vec{z}) \not \equiv 0.$

For (\ref{genNLS}), 
we find
\begin{equation} 
  \label{genNLScase1}
  \begin{cases}
    \alpha_1 = \alpha_2 = \alpha_3 = \alpha_4 = -1, \\
    u_0(x,t) = -2 g_x^2(x,t) (1+\beta)^{-1} \bar{u}_0^{-1}(x,t), \\
    v_0(x,t) = -2 g_x^2(x,t) (1+\beta)^{-1} \bar{v}_0^{-1}(x,t), \\
  \end{cases}
\end{equation}
where $\bar{u}_0(x,t)$ and $\bar{v}_0(x,t)$ are arbitrary functions.  

If we do not restrict $\alpha_1,\dotsc,\alpha_4<0,$ then there are 
contradictions with the assumption $u_{i,0}(\vec{z}) \not \equiv 0$ 
for all but two possible dominant behaviors,
\begin{equation} 
  \label{genNLScase2}
  \begin{cases}
    \alpha_1 = \alpha_2 = -1, \\
    \alpha_3 \ge 3, \\
    \alpha_4 \ge 3, 
  \end{cases} 
  \qquad \text{and} \qquad 
  \begin{cases}
    \alpha_1 \ge 3, \\
    \alpha_2 \ge 3,  \\
    \alpha_3 = \alpha_4 = -1. 
  \end{cases}
\end{equation}

% % % % % % % % % % % % % % % % % % % % % % % % % % % % % % % % 
\subsection{Algorithm to determine the resonances}
\label{algo:resonancesSection}

%  %  %  %  %  %  %  %  %  %  %  %  %  %  %  %  %  %  %  %  %  %
\step[Construct matrix $Q_r$]

Substitute 
\begin{equation}
  u_i(\vec{z}) = u_{i,0}(\vec{z}) g^{\alpha_i}(\vec{z}) 
    + u_{i,r}(\vec{z}) g^{\alpha_i + r}(\vec{z})
\end{equation}
into (\ref{painleveSystem}).  
Then, the $(i,j)$-th entry of the $M\times M$ matrix $Q_r$ is 
the coefficients of the linear terms in $u_{j,r}(\vec{z})$ 
of the leading terms in equation $F_i=0.$ 

%  %  %  %  %  %  %  %  %  %  %  %  %  %  %  %  %  %  %  %  %  %
\step[Find the roots of $\det(Q_r)$]

The resonances are the solutions of $\det(Q_r) = 0.$  
If any of these solutions (in a particular branch) is non-integer, 
then that branch of the algorithm terminates since it implies that 
some solutions of (\ref{painleveSystem}) have movable algebraic branch point.  
If any of the resonances are rational, then a change of variables in 
(\ref{painleveSystem}) may remove the algebraic branch point.  
Such changes are not carried out automatically.  

For branch (\ref{genNLScase1}), 
\begin{gather}
  \det(Q_r) = 
  (r-4)(r-3)^2 
  \left\{r^2(1+\beta)-3r(1+\beta)-4(1-\beta)\right\}r^2(r+1) \notag\\  
    \label{genNLSdetQr}
    \qquad{} \times(1+\beta)^5\bar{u}_0^5(x,t)\bar{v}_0^5(x,t)g_x^8(x,t).
\end{gather}
Since the roots of (\ref{genNLSdetQr}) for $r$ depend 
on the constant parameter $\beta,$ 
we must choose values of $\beta$ so that all the solutions 
are integers before proceeding.  
For $\beta = 1,$ the resonances are  
$r_1 = -1, r_2 = r_3 = r_4 = 0, r_5 = r_6 = r_7 = 3, r_8 = 4.$ 

While taking $\beta = 0$ also yields all integer resonances, 
it violates the assumption that all the parameters in 
(\ref{painleveSystem}) are nonzero.  
Allowing the parameters in (\ref{painleveSystem}) to be zero 
could cause a false balance in Algorithm~\ref{algo:dominantBehavior}.  
Thus, (\ref{painleveSystem}) with $\beta = 0$ should be treated separately.
In this example however, 
setting $\beta = 0$ does not affect the dominant behavior 
and the resonances are $r_1 = r_2 = -1, r_3 = r_4 = 0, r_5 = r_6 = 3,$ and 
$r_7 = r_8 = 4.$  

Although taking $\beta = 25/7$ leads to rational resonances at 
$r_4 = r_5 = 3/2,$ they are not easily resolved by a change 
of variables in (\ref{genNLS}).
The branches with dominant behavior, 
$\alpha_1 = \alpha_2 = -1, \alpha_3\ge\alpha_4 \ge 3,$
have resonances 
$r_1 = -\alpha_3-1, r_2 = -\alpha_3+2, r_3 = -\alpha_4-1, 
r_4 = -\alpha_4+2, r_5 = -1, r_6 = 0, r_7 = 3$ and $r_8 = 4.$
Since $r_1,r_2 < -1$ when $\alpha_3=\alpha_4=3,$ 
$r_1,r_2,r_3 < -1$ when $\alpha_3 > \alpha_4=3,$ and 
$r_1,\dotsc,r_4 < -1$ when $\alpha_3,\alpha_4 > 3,$ 
these are not principal branches and 
should be investigated using 
the perturbative Painlev\'e approach~\cite{CFP93}. 

% % % % % % % % % % % % % % % % % % % % % % % % % % % % % % % % 
\subsection{Algorithm to determine the constants of integration 
  and check compatibility conditions}

%  %  %  %  %  %  %  %  %  %  %  %  %  %  %  %  %  %  %  %  %  %
\step[Generate the system for the coefficients of the Laurent series 
at level $k$]

Substitute
\begin{equation}
    u_i(\vec{z}) = g^{\alpha_i}(\vec{z}) 
    \sum_{k = 0}^{r_m} u_{i,k}(\vec{z}) g^k(\vec{z})
\end{equation}
into (\ref{painleveSystem}) and multiply $F_i$ by $g^{-\gamma_i}(\vec{z}),$ 
where $\gamma_i$ is the lowest exponent of $g(\vec{z})$ in $F_i.$  
The equations for determining the coefficients of the Laurent series at 
level $k$ then arise by equating to zero the coefficients of $g^k(\vec{z}).$  
These equations, at level $k,$ are linear in $u_{i,k}(\vec{z})$ and 
depend only on $u_{i,j}(\vec{z})$ and $g(\vec{z})$ (and their derivatives)
for $1 \le i \le M$ and $0 \le j < k.$
Thus, the system can be written as 
\begin{equation}
  \label{painLinSystem}
    Q_k \vec{u}_k 
    = \vec{G}_k(\vec{u}_0,\vec{u}_1,\dotsc,\vec{u}_{k-1},g,\vec{z}),
\end{equation}
where $\vec{u}_k = (u_{1,k}(\vec{z}), \dotsc, u_{M,k}(\vec{z}))^T.$ 

\step[Solve the linear system for the coefficients of the Laurent series]

If the rank of $Q_k$ equals the rank of the augmented matrix 
$(Q_k|\vec{G}_k),$ 
solve (\ref{painLinSystem}) for the coefficients of the Laurent series.  
If $k=r_j,$ check that $\rank Q_k = M-s_j,$ 
where $s_j$ is the algebraic multiplicity of the resonance $r_j$ in 
$\det(Q_r) = 0.$  

If $\rank Q_k \neq \rank (Q_k|\vec{G}_k),$ 
Gauss reduce the augmented matrix $(Q_k|\vec{G}_k)$ to determine 
the compatibility condition.  
If all the compatibility conditions can be resolved by restricting the 
coefficients parameterizing (\ref{painleveSystem}), 
then (\ref{painleveSystem}) has the Painlev\'e property for those 
specific values.  
If any of the compatibility conditions cannot be resolved by restricting the 
coefficients parameterizing (\ref{painleveSystem}), 
then the Laurent series solution for this branch has 
a movable logarithmic branch point and the algorithm terminates.  

For (\ref{genNLS}) with $\beta = 1,$ 
the principal branch
\begin{equation}
  \begin{cases}
    \alpha_1 = \alpha_2 = \alpha_3 = \alpha_4 = -1, \\
    u_0(x,t) = 
    -\bar{u}_0^{-1}(x,t) \{ v_0(x,t)\bar{v}_0(x,t) - 2g_x^2(x,t)\}, \\
    \bar{u}_0(x,t), v_0(x,t), \bar{v}_0(x,t) \text{ arbitrary}, \\
    r_1 = -1, r_2 = r_3 = r_4 = 0, r_5 = r_6 = r_7 = 3, r_8 = 4,
  \end{cases}
\end{equation}
has three compatibility conditions at level $k = r_5 = r_6 = r_7 = 3.$  
These compatibility conditions require that 
$a_x(x,t) = \bar{a}_x(x,t) = b_x(x,t) = \bar{b}_x(x,t)$  
and $c_x(x,t) = d_x(x,t) = 0.$  
\vfill
\newpage

At level $k = r_8 = 4,$ 
the compatibility condition requires 
$d(t) = \bar{c}(t)$ and 
\begin{gather}
  \{(a-\bar{a})^2+2i(a_x-\bar{a}_x)h'(t)\}(2+v_0\bar{v}_0)
  - \{(b-\bar{b})^2+2i(b_x-\bar{b}_x)h'(t)\}v_0\bar{v}_0 \notag\\
  \label{NLScomplexconstraints} \qquad{}
  +2i(a-\bar{a}+b-\bar{b})(v_0(\bar{v}_0)_t+(v_0)_t\bar{v}_0)
  + i(a_t - \bar{a}_t - b_t + \bar{b}_t)v_0\bar{v}_0 \\
  \qquad{}
  +2i(a_t - \bar{a}_t)
  - (a_{xx} + \bar{a}_{xx} - b_{xx} - \bar{b}_{xx})v_0\bar{v_0} 
  - 2 a_{xx} + 6 \bar{a}_{xx} \equiv 0,\notag
\end{gather}
where we have taken $g(x,t) = x - h(t).$  
Careful inspection of (\ref{NLScomplexconstraints}) 
reveals that $a(x,t) = b(x,t).$  
Setting $a(x,t) = b(x,t) = r(x,t) + i s(x,t),$ 
where $r(x,t)$ and $s(x,t)$ are arbitrary real functions, 
(\ref{NLScomplexconstraints}) becomes
\begin{equation}
  2 s^2(x,t) + s_t(x,t) + 2h'(t)s_x(x,t) - 
  r_{xx}(x,t) + 2is_{xx}(x,t) \equiv 0.
\end{equation}
Since $h'(t)$ is arbitrary, 
it follows that $s_x(x,t) = 0.$ 
Thus, $r_{xx}(x,t) = 2 s^2(t) + s'(t)$ and upon integration 
\begin{equation}
  r(x,t) = \frac{1}{2}\{ 2 s^2(t) + s'(t) \}x^2 + r_1(t) x + r_2(t),
\end{equation}
where $r_1(t)$ and $r_2(t)$ are arbitrary functions.

Therefore, the generalized coupled NLS equations,
\begin{equation*}
  \begin{gathered}
  iu_t + u_{xx} + (|u|^2 + |v|^2)u 
    + \left\{ \{s^2(t) + \tfrac{1}{2}s'(t)\}x^2  
    + r_1(t) x + r_2(t) + i s(t)\right\}u + c(t) v = 0, \\
  iv_t + v_{xx} + (|u|^2 + |v|^2)v 
    + \left\{ \{s^2(t) + \tfrac{1}{2}s'(t)\}x^2 
    + r_1(t) x + r_2(t) + i s(t)\right\}v + \bar{c}(t) u = 0, 
  \end{gathered}
\end{equation*}
passes the Painlev\'e test, 
where $r_1(t),r_2(t),$ and $s(t)$ are arbitrary real functions
and $c(t)$ is an arbitrary complex function.

When $\beta = 0,$ 
the two compatibility conditions at level $k = r_5 = r_6 = 3$ 
require that $c(x,t) = d(x,t) = \bar{c}(x,t) = \bar{d}(x,t) = 0.$  
Similarly, the compatibility conditions at level $k = r_7 = r_8 = 4,$ 
require that 
\begin{equation}
  a(x,t) = \{s^2(t) - \tfrac{1}{2}s'(t)\}x^2  + r_1(t) x + r_2(t) + i s(t),
\end{equation} 
where $r_1(t), r_2(t)$ and $s(t)$ are arbitrary real functions.  
Therefore, 
\begin{equation}
iu_t + u_{xx} + |u|^2u 
    + \left\{ \{s^2(t) - \tfrac{1}{2}s'(t)\}x^2  
    + r_1(t) x + r_2(t) + i s(t)\right\}u = 0,
\end{equation}
passes the Painlev\'e test, 
a fact confirmed in~\cite{Ablowitz91}.

% \vfill
% \newpage
%%%%%%%%%%%%%%%%%%%%%%%%%%%%%%%%%%%%%%%%%%%%%%%%%%%%%%%%%%%%%%%%
\section{Additional Examples}
\label{sec:additionalExamples}

%  %  %  %  %  %  %  %  %  %  %  %  %  %  %  %  %  %  %  %  %  %  %
%  %  %  %  %  %  %  %  %  %  %  %  %  %  %  %  %  %  %  %  %  %  %
% 
% @@@ begin change WH 01/03/2006 ODE example added
% 
\subsection{A peculiar ODE}

%
% @@@ begin change DB 1/4/2006, reformulated
%
Consider the ODE~\cite{RandW86} 
\begin{equation}
  \label{randwinternitzode}
  u^2 u''' - 3 (u')^3 = 0.
\end{equation}
Substituting (\ref{dominantBehavior}) into (\ref{randwinternitzode}) 
gives $\alpha (\alpha + 2 ) (2 \alpha - 1) \chi^3 g(z)^{3(\alpha - 1)} = 0.$  
So, both the terms in (\ref{randwinternitzode}) have the same 
leading exponent, $3(\alpha - 1).$ 
Using the procedure in Section \ref{algo:dominantBehavior}, in Step 5 the 
software automatically runs the test for the default values 
$\alpha = -3, -2,$ and $ -1.$
The choices $\alpha = -1 $ and $-3$ are incompatible with the assumption
$u_0 \neq 0.$
The leading term vanishes for $\alpha = -2$ and $u_0$ is arbitrary.  
Substituting $u(z) = u_0 \, g^{-2}(z) + u_{1,r} \, g^{r-2}(z),$ 
we find that $r_1 = -1, r_2 = 0,$ and $r_3 = 10.$  
Thus, the Laurent series solution of (\ref{randwinternitzode}) is
\begin{equation}
  u(z) = u_0 (z - z_0)^{-2} + u_{10} (z - z_0)^8 + \dotsb,
\end{equation}
where $z_0,u_0$ and $u_{10}$ are arbitrary constants.  
Hence, (\ref{randwinternitzode}) passes the Painlev\'e test.
%
% @@@ end change DB 1/4/2006
%

%Consider the following ODE \cite{RandW86} for $u(z):$
%\begin{equation}
%\label{randwinternitzode}
%u^2 u^{\prime\prime\prime} - 3 u^{\prime 3} = 0.
%\end{equation}
%% 
%% for $u(z).$
%Upon substitution of $u(z) \!=\! u_1(z) \!=\! \chi_1 g^{\alpha_1}$ into 
%(\ref{randwinternitzode}) one obtains 
%$-\alpha_1 (\alpha_1 + 2 ) (2 \alpha_1 - 1) \chi_1^3 g^{3(\alpha_1 -1)}.$
%So, both terms in (\ref{randwinternitzode}) have the same leading exponent, 
%namely $3(\alpha_1 -1).$ 
%Hence, $\alpha_1$ remains undetermined and the software automatically runs 
%the test for the default values $\alpha_1 = -3, -2,$ and $ -1.$
%Note that the dominant term vanishes for $\alpha_1 = -2,$ which is the only 
%case for which $u_{1,0} \ne 0.$ 
%As a matter of fact, $u_{1,0}$ is arbitrary for $\alpha_1 = -2.$
%% 
%Upon substitution of $u_1(z) = u_{1,0} \, g^{-2} + u_{1,r} \, g^{r-2}$ 
%with arbitrary $u_{1,0},$ one finds $r_1 = -1, r_2 = 0,$ and $r_3 = 10.$
%% 
%Next, upon substitution of 
%$u_1(z) = g^{-2} \sum_{k=0}^{10} u_{1,k} \, g^{k}$ one determines that 
%$u_{1,k} = 0$ for $k=1,\cdots, 9$ and that $u_{1,10}$ is arbitrary.
%Hence, (\ref{randwinternitzode}) passes the Painlev\'e test.
% 
% @@@ end change WH 01/03/2006
% 

\subsection{The sine-Gordon equation}

Consider the sine-Gordon equation~\cite{Ablowitz91},
\begin{equation}
  \label{sineGordon}
  u_{tt} + u_{xx} = \sin{u}.
\end{equation}
Using the transformation $v(x,t) = e^{i\,u(x,t)},$ 
we obtain a polynomial differential equation 
\begin{align}
%  \left(\log{v}\right)_{tt} + \left(\log{v}\right)_{xx} & = 
%    \frac{e^{iu} - e^{-iu}}{2}, \\
%  \left(\frac{v_t}{v}\right)_t + \left(\frac{v_x}{v}\right)_x & = 
%    \frac{v - v^{-1}}{2}, \\
%  \frac{vv_{tt} - v_t^2 }{v^2} + \frac{vv_{xx} - v_x^2 }{v^2} & = 
%    \frac{v - v^{-1}}{2}, \\ 
  \label{polySineGordon}
  vv_{tt} + vv_{xx} - v_t^2 - v_x^2 & = \frac{1}{2}v(v^2 - 1).
\end{align}
The dominant behavior of (\ref{polySineGordon}) is 
$v(x,t) \sim 4(g_x^2(x,t) + g_t^2(x,t))g^{-2}(x,t),$  with 
resonances $r_1 = -1$ and $r_2 = 2.$  
The Laurent series solution of (\ref{polySineGordon}) is 
\begin{equation}
  v = 4(g_x^2 + g_t^2)g^{-2} - 4(g_{xx} + g_{tt})g^{-1} + v_2 + \dotsb,
\end{equation}
where $g(x,t)$ and $v_2(x,t)$ are arbitrary functions.
The sine-Gordon equation passes the Painlev\'e test 
and is indeed completely integrable~\cite{Ablowitz91}.  

%  %  %  %  %  %  %  %  %  %  %  %  %  %  %  %  %  %  %  %  %  %  %
%\subsection{Liouville equation}

%Similarly, the \emph{Liouville equation}~\cite{Ablowitz91},
%\begin{equation}
%  \Delta u = \exp(u),
%\end{equation}
%becomes the polynomial PDE 
%\begin{equation}
%    vv_{tt} + vv_{xx} - v_t^2 - v_x^2 = v^3,
%\end{equation}
%with the transformation $v(x,t) = e^{u(x,t)}.$  
%The dominant behavior of (\ref{polySineGordon}) is 
%$v(x,t) \sim 2(g_x^2(x,t) + g_t^2(x,t))g^{-2}(x,t),$  with 
%resonances at levels $r_1 = -1$ and $r_2 = 2.$  
%The general solution of (\ref{polySineGordon}) is 
%begin{equation}
%  v = 2(g_x^2 + g_t^2)g^{-2} - 2(g_{xx} + g_{tt})g^{-1} + v_2 + \dotsb,
%\end{equation}
%where $g$ and $v_2$ are arbitrary functions of $x$ and $t.$  
%Hence, we conclude that the Liouville equation passes the Painlev\'e test.  

\subsection{The cylindrical Korteweg-de Vries equation}

Consider the generalized KdV equation,
\begin{equation}
  \label{ckdv}
  u_t + 6u u_x + u_{3x} + a(t) u = 0,
\end{equation}
where $a(t)$ is an arbitrary function parameterizing the equation.  
The dominant behavior of (\ref{ckdv}) is 
$u(x,t) \sim -2 g_x^2 (x,t) g^{-2}(x,t),$ 
with resonances $r_1=-1, r_2=4$ and $r_3=6.$  
At level $k=r_3=6,$ we obtain the compatibility condition
\begin{equation}
  \frac{2 a(t)^2 + a'(t)}{6 g_x(x,t)} = 0.
\end{equation}
So, (\ref{ckdv}) passes the Painlev\'e test if $a(t) = \tfrac{1}{2t}.$ 
In this case, (\ref{ckdv}) reduces to the cylindrical KdV, which is 
completely integrable as confirmed by other analyses~\cite{Ablowitz91}. 

%  %  %  %  %  %  %  %  %  %  %  %  %  %  %  %  %  %  %  %  %  %  %
\subsection{A fifth-order generalized Korteweg-de Vries equation}

Consider the generalized fifth-order KdV equation,
\begin{equation}
  \label{gen5kdv}
  u_t + a u_xu_{xx} + b uu_{3x} + c u^2u_x + u_{5x} = 0,
\end{equation} 
with constant parameters $a, b,$ and $c.$
The dominant behavior of (\ref{gen5kdv}) is
\begin{equation}
  u(x,t) \sim -\frac{3g_x^2(x,t)}{c} \left\{ (a+2b) \pm 
    \sqrt{a^2 + 4 ab + 4b^2-40c}\right\} g^{-2}(x,t).
\end{equation}
The resonances are the roots of 
\begin{gather}
  \det(Q_r) = -c (r-6)(r+1) \Big(3 \sqrt{(a+2b)^2 - 40c}(2a-b(r-4)) \notag \\
  \label{gen5kdvRoots}
    \qquad{} 
     - 6 (a+2b)^2 + 240 c + (3b(a+2b)-86c)r + 15 cr^2 - c r^3 \Big) g_x^5.
\end{gather} 

Determining what values of $a,b,$ and $c$ that lead to integer roots of 
(\ref{gen5kdvRoots}) is difficult by hand or with a computer.  
An investigation of the scaling properties of (\ref{gen5kdv}) reveals that 
only the ratios $a/b$ and $c/b^2$ are important.  
Let us consider the well-known special cases.  

If we take $a = b$ and $5c = b^2,$ then (\ref{gen5kdv}) passes the 
Painlev\'e test with resonances
$r_1 = -2, r_2 = -1, r_3 = 5, r_4 = 6, r_5 = 12$ and 
$r_1 = -1, r_2 = 2, r_3 = 3, r_4 = 6, r_5 = 10.$  
Taking $b = 5,$ equation (\ref{gen5kdv}) becomes the completely 
integrable equation 
\begin{equation}
  u_t + 5 u_xu_{xx} + 5 uu_{3x} + 5 u^2u_x + u_{5x} = 0,
\end{equation}
due to Sawada and Kotera~\cite{Sawada74} and Caudrey et al.~\cite{Caudrey76}.  

If we take $a = 2 b$ and $10 c = 3 b^2,$ then (\ref{gen5kdv}) passes the 
Painlev\'e test with resonances 
$r_1 = -3, r_2 = -1, r_3 = 6, r_4 = 8, r_5 = 10$ and 
$r_1 = -1, r_2 = 2, r_3 = 5, r_4 = 6, r_5 = 8.$  
For $b = 10,$ equation (\ref{gen5kdv}) is a member of the 
completely integrable KdV hierarchy 
\begin{equation}
  \label{laxEqn}
  u_t + 10 uu_{3x} + 20 u_xu_{xx} + 30 u^2u_x + u_{5x} = 0,
\end{equation}
due to Lax~\cite{Lax68}.  

If we take $2 a = 5 b$ and $5 c = b^2,$ then (\ref{gen5kdv}) passes the 
Painlev\'e test with resonances 
$r_1 = -7, r_2 = -1, r_3 = 6, r_4 = 10, r_5 = 12$ and 
$r_1 = -1, r_2 = 3, r_3 = 5, r_4 = 6, r_5 = 7.$  
When $b = 10,$ equation (\ref{gen5kdv}) is the Kaup-Kupershmidt 
equation~\cite{Fordy80,Hirota80}, 
\begin{equation}
  u_t + 10 uu_{3x} + 20 u^2u_x + 25 u_xu_{xx} + u_{5x} = 0,
\end{equation}
which is also known to be completely integrable.

While there are many other values for $a,b,$ and $c,$ 
for which (\ref{gen5kdvRoots}) only has integer roots, 
but compatibility conditions prevent (\ref{gen5kdv}) 
from having the Painlev\'e property.  
For instance, when $a = 2b$ and $5 c = 2 b^2,$ 
the resonances are $r_1 = -1, r_2 = 0, r_3 = 6, r_4 = 7, r_5 = 8.$  
At level $k = r_2 = 0,$ we are forced to take 
$u_0(x,t) = - 30 g_x^2(x,t)/b,$ so the Laurent series solution is 
not the general solution and (\ref{gen5kdv}) fails the Painlev\'e test.  
Similarly, when $7 a = 19 b$ and $49 c = 9 b^2,$ 
we have resonances $r_1 = -1, r_2 = 3$ and $r_3 = r_4 = r_5 = 6,$ 
so the Laurent series solution is not the general solution and, 
again (\ref{gen5kdv}) fails the Painlev\'e test.  

%%%%%%%%%%%%%%%%%%%%%%%%%%%%%%%%%%%%%%%%%%%%%%%%%%%%%%%%%%%%%%%%
\section{Brief Review of Symbolic Algorithms and Software}
\label{sec:compare}

There is a variety of methods for testing nonlinear ODEs and PDEs 
for the Painlev\'e property.  
While the WTC algorithm discussed in this paper is 
the most common method used in Painlev\'e analysis, 
it is not appropriate in all cases.  
For instance, there are numerous completely integrable differential 
equations which have algebraic branching in their series solutions; 
a property that is allowed by the so-called ``weak'' Painlev\'e test 
% 
% @@@ begin change WH 01/03/2006 added reference to Goriely's book
% 
(see~\cite{Goriely01,Ramani82,Ramani89}).  
% 
% @@@ end change WH 01/03/2006
% 
A more thorough approach for testing differential equations with 
branch points is the poly-Painlev\'e test (see~\cite{Kruskal92,Kruskal97}).  
The perturbative Painlev\'e test~\cite{CFP93} was developed to check the 
compatibility conditions of negative resonances other than $r=-1.$ 

For testing ODEs, there are several implementations:  
\texttt{ODEPAINLEVE} developed by Rand and Winternitz~\cite{RandW86} 
in \emph{Macsyma} is restricted to scalar differential equations;
\texttt{PTEST.RED} by Renner in \emph{Reduce}~\cite{Renner92}; and,
a \emph{Reduce} package by Scheen~\cite{Scheen97} which implements
both the traditional and the perturbative Painlev\'e tests.
For testing PDEs, there are a few implementations.  
The package \texttt{PAINMATH.M} by Hereman et al.~\cite{Hereman98} 
is unable to find all the dominant behaviors in systems with undetermined 
$\alpha_i$ and is limited to two independent variables.   
%A new method by Xie and Chen~\cite{Xie03} developed for \emph{Maple} 
%does not appear to be completely automated.  

Only the \emph{Maple} package \texttt{PDEPtest} by 
Xu and Li~\cite{Xu03,Xu04,Xu05} 
is comparable to our package \texttt{PainleveTest.m}~\cite{painlevecode01}.  
The package \texttt{PDEPtest} was written after our package and allows the 
testing of systems of PDEs (but not ODEs) parameterized by arbitrary 
functions using either the traditional WTC algorithm or the simplification 
% 
% @@@ begin change WH 01/03/2006 purposed --> proposed
% 
proposed by Kruskal (see Section~\ref{sec:algorithm}).  
% 
% @@@ end change WH 01/03/2006
% 
While \texttt{PDEPtest} can find all the dominant behaviors in some 
systems with undetermined $\alpha_i$ (such as the Hirota-Satsuma system), 
it fails to find the dominant behaviors for systems in which more than 
one $\alpha_i$ is undetermined
(such as the NLS equation, $iu_t + u_{zz} + 2u|u|^2 = 0,$ 
which is completely integrable~\cite{Ablowitz91}).  
Furthermore, 
\texttt{PDEPtest} requires that all the $\alpha_i$ are negative, 
a weakness of the implementation, 
since it is standard to allow some positive exponents 
(see equation (2.4) in \cite{Ramani89} with leading exponents $-1$ and $1).$

%%%%%%%%%%%%%%%%%%%%%%%%%%%%%%%%%%%%%%%%%%%%%%%%%%%%%%%%%%%%%%%%
\section{Using the Software Package {\ttfamily PainleveTest.m}}
\label{sec:usage}

The package \texttt{PainleveTest.m} has been tested on both PCs and 
UNIX work stations with \textit{Mathematica} 
versions 3.0, 4.0, 4.1, 5.0, 5.1, and 6.0 
% 
% @@@ begin change WH 01/03/2006 two dozen instead of half a dozen ???
% 
using a test set of over 50 PDEs and two dozen ODEs.  
% 
% @@@ end change WH 01/03/2006
% 
The Backus-Naur form of the function is 
\begin{tabbing}
$\qquad$ \= 
$\langle Main\, Function \rangle \quad \to \quad$ \=
$\texttt{PainleveTest}[\langle Equations \rangle, 
\langle Functions \rangle,$ \+ \+ \\
$\qquad \langle Variables \rangle, \langle Options \rangle]$ \- \\ 
$\langle Options \rangle \quad \to \quad $ \= 
$\verb|Verbose|\rightarrow \langle Boolean \rangle \; | $ \+ \\
$\verb+KruskalSimplification+\rightarrow \langle Variable \rangle \; |$ \\
$\verb|DominantBehaviorMin|\rightarrow \langle 
Negative~Integer \rangle \; |$ \\
$\verb|DominantBehaviorMax|\rightarrow \langle Integer \rangle \; | $\\
$\verb|DominantBehavior| \rightarrow 
\langle List\, of\, Rules \rangle \; |$ \\
$\verb|DominantBehaviorConstraints| \rightarrow 
\langle List\, of\, Constraints \rangle \; |$ \\
$\verb|DominantBehaviorVerbose|\rightarrow \langle Range \rangle \; |$ \\
$\verb|ResonancesVerbose|\rightarrow \langle Range \rangle \; | $ \\
$\verb|ConstantsOfIntegrationVerbose|\rightarrow \langle Range \rangle$ \- \\
$\langle Bool \rangle \quad \to \quad \mathtt{True} \; | \; \mathtt{False}$ \\
$\langle Range \rangle \quad \to \quad 0 \; | \; 1 \; | \; 2  \; | \; 3$ \\
$\langle List\, of\, Rules \rangle \quad \to \quad \mathtt{
\{\{ alpha[1] \rightarrow} \langle Integer \rangle\mathtt{, 
  alpha[2] \rightarrow}\langle Integer \rangle \mathtt{,... \},... \} }$ \\
$\langle List\, of\, Constraints \rangle \quad 
\to \quad \mathtt{ \{ alpha[1] == alpha[2],... \}}$ \\
\end{tabbing}
The output of the function is 
\begin{gather*}
  \Big\{ \big\{ \{ \text{Dominant behavior} \}, \{ \text{Resonances} \}, \\
  \phantom{\Big\{ \big\{} \{ \{ \text{Laurent series coefficients} \}, 
    \{ \text{Compatibility conditions}\} \big\}, 
    \dotsc \Big\}
\end{gather*}
% 
% @@@ begin change WH 01/03/2006 c_1(t), c_2(t) instead of c_1(x,t), c_2(x,t)
% 
If using a PC, place the package \texttt{PainleveTest.m} in a directory, 
say myDirectory on drive C.  
Start a \textit{Mathematica} notebook session and execute the commands:
\begin{verbatim}
In[1] = SetDirectory["c:\\myDirectory"];  (* Specify the directory *)

In[2] = Get["PainleveTest.m"]             (* Read in the package *)

In[3] = PainleveTest[                     (* Test the KdV equation *)
         {D[u[x,t],t]+6*u[x,t]*D[u[x,t],x]+D[u[x,t],{x,3}] == 0}, 
         u[x, t], {x,t}, KruskalSimplification -> x]
Out[3] = 
\end{verbatim}
\begin{gather*}
 \Big\{ \big\{ \{ {{\alpha_1}\rightarrow {-2}}\} ,
   \{ {r\rightarrow {-1}},{r\rightarrow 4},{r\rightarrow 6}\} , \\
   \quad \{ \{ {{u_{1,0}}\rightarrow {-2}},{{u_{1,1}}\rightarrow 0},
     {{u_{1,2}}\rightarrow {\frac{h'(t)}{6}}},{{u_{1,3}}\rightarrow 0}, \\
   \qquad
     {{u_{1,4}}\rightarrow {C_1(t)}},
     {{u_{1,5}}\rightarrow {\frac{h''(t)}{36}}},
     {{u_{1,6}}\rightarrow {C_2(t)}}\} ,
    \{ \} \} \big\} \Big\}
\end{gather*}
% 
% @@@ end change WH 01/03/2006 
% 
The option \verb|KruskalSimplification -> x| allows one to use 
$g(x,t) = x - h(t)$ in the calculation of the constants of integration 
and in checking the compatibility conditions.  
% 
% @@@ begin change WH 01/03/2006 alpha_2 instead of alpha_1 in first case
% 
\begin{verbatim}
In[4] = PainleveTest[  (* Eq. (2.4) in Ramani et al. [32] *)
         {D[x[z], z] == x[z]*(a - x[z] - y[z]),
          D[y[z], z] == y[z]*(x[z] - 1)}, 
         {x[z], y[z]}, {z}, DominantBehaviorMax -> 1 ]
Out[4] = 
\end{verbatim}
% 
% @@@ begin change WH 01/03/2006 g'(z) ----> 1 and C_1(x) ---> C_1 
% 
\begin{gather*}
 \Big\{ \big\{ \{ {{\alpha_1}\rightarrow {-1},{\alpha_2}\rightarrow {-1}}\} ,
   \{ {r\rightarrow {-1}},{r\rightarrow 2}\} , 
    \{ \{ {{u_{1,0}}\rightarrow {-1}},\dotsc
     \} , 
    \{ a + 1 = 0 \} \} \big\}, \\
    \big\{ \{ \alpha_1 \rightarrow -1, \alpha_2 \rightarrow 1 \},
      \{ r \rightarrow -1, \rightarrow 0\}, 
      \{\{ u_{1,0} \rightarrow 1, u_{2,0} \rightarrow C_1 \}, \{ \} \}
    \big\}\Big\}
\end{gather*}
In this example, if the \verb|DominantBehaviorMax| option was not used, 
we would wrongly conclude that the system only passes the Painlev\'e test 
when $a = -1.$
However, by allowing positive $\alpha_i,$ we find the second branch 
$\alpha_1 = -1$ and $\alpha_2 = 1,$ 
for which the system passes the Painlev\'e test without restricting the 
value of the parameter $a.$
% 
% @@@ end change WH 01/03/2006 
% 
Alternatively, executing
\begin{verbatim}
In[5] = PainleveTest[{ D[x[z], z] == x[z]*(a - x[z] - y[z]),
          D[y[z], z] == y[z]*(x[z] - 1)}, {x[z], y[z]}, {z}, 
          DominantBehavior -> {{alpha[1] -> -1, alpha[2] -> 1}} ]
\end{verbatim}
would only test the branch with $\alpha_1 = -1$ and $\alpha_2 = 1.$
For an example of the option \texttt{DominantBehaviorConstraints},
% 
% @@@ begin change WH 01/03/2006 reformulated
% 
see Step 4 of Algorithm~\ref{algo:dominantBehavior}.
% 
% @@@ end change WH 01/03/2006 
% 

The option \verb|Verbose -> True| gives a brief trace of the calculations 
in each of the three steps of the algorithm.  
The options 
\texttt{DominantBehaviorVerbose}, \texttt{ResonancesVerbose}, 
and \texttt{ConstantsOfIntegrationVerbose} allow for a more detailed 
trace of the calculation.  
For instance, \verb|DominantBehaviorVerbose -> 1| 
would show the result of substituting the ansatz, 
the exponents before and after removing non-dominant powers, etc.   
While \verb|DominantBehaviorVerbose -> 3| shows the result of nearly 
every line of code in the package, allowing the user to check the results 
in the trickiest cases. 

%%%%%%%%%%%%%%%%%%%%%%%%%%%%%%%%%%%%%%%%%%%%%%%%%%%%%%%%%%%%%%%%
\section{Discussion and Conclusions}
\label{sec:conclusion}

Our software package \texttt{PainleveTest.m} is applicable to polynomial 
systems of nonlinear ODEs and PDEs.  
While the Painlev\'e test does not guarantee complete integrability, 
it helps in identifying candidate differential equations for complete 
integrability in a straightforward manner.  
For differential equations with parameters 
(including arbitrary functions of the independent variables), 
our software allows the user to determine the conditions under 
which the differential equations may possess the Painlev\'e property.  
Therefore, by finding the compatibility conditions,
classes of parameterized differential equations can be analyzed and 
candidates for complete integrability can be identified.  

The difficulty in completely automating the Painlev\'e test lies 
in determining the dominant behaviors of the Laurent series solutions;   
specifically, determining \emph{all} the valid dominant behaviors when 
one or more of the $\alpha_i$ are undetermined.  
While there are other implementations for the Painlev\'e test, 
ours is currently the only implementation in \emph{Mathematica} which 
allows the testing of polynomial systems of nonlinear PDEs with no 
limitations on the number of differential equations or the number of 
independent variables (except where limited by memory).  

%%%%%%%%%%%%%%%%%%%%%%%%%%%%%%%%%%%%%%%%%%%%%%%%%%%%%%%%%%%%%%%%

% 
\label{lastpage}
\end{document}